# Custom edge-element FEM solver and its application to eddy-current simulation of realistic 2M-element human brain phantom

Wuliang Yin[1,2], Mingyang Lu[2], Jiawe Tang[2], Qian Zhao[3], Zhijie Zhang[1], Kai Li[4], Yan Han[5], Anthony Peyton[2],

[1] School of Instrument and Electronics, North University of China, Taiyuan, Shanxi, China

[2] School of Electrical and Electronic Engineering, University of Manchester, Manchester, UK

[3] College of Engineering, Qufu Normal University, Shandong, China

[4] School of Instrument and Electronics, North University of China, Taiyuan, Shanxi, China

[5] School of Information and Communication Engineering, North University of China, Taiyuan, China

Corresponding author:

Mingyang Lu

School of Electrical and Electronic Engineering, University of Manchester, Manchester, UK.

M1 3WE

Tel: +44 (0) 161 306 2885.

Email: Mingyang.lu@manchester.ac.uk

Running title: Applications of a custom FEM solver

Conflicts of Interest: None.

Grant information: None.

# ABSTRACT

Extensive research papers of three-dimensional computational techniques are widely used for the investigation of human brain pathophysiology. Eddy current analysing could provide an indication of conductivity change within a biological body. A significant obstacle to current trend analyses is the development of a numerically stable and efficiency finite element scheme which performs well at low frequency and does not require a large number of degrees of freedom. Here, a custom finite element method (FEM) solver based on edge elements is proposed using the weakly coupled theory, which separates the solution into two steps. First, the background field (the magnetic vector potential on each edge) is calculated and stored. Then, the electric scalar potential on each node is obtained by FEM based on Galerkin formulations. Consequently, the electric field, eddy current distribution in the object can be obtained. This solver is more efficient than typical commercial solvers since it reduces the vector eddy current equation to a scalar one, and reduces the meshing domain to just the eddy current region. It can therefore tackle complex eddy current calculations for models with much larger numbers of elements, such as those encountered in eddy current computation in biological tissues. An example is presented with a realistic human brain mesh of 2 million elements. In addition, with this solver, the equivalent magnetic field induced from the excitation coil is applied and therefore there is no need to mesh the excitation coil. In combination, these have significantly increased the efficiency of the solver.

# Introduction

In electromagnetic computation, the finite-element method (FEM) [Kinouchi et al., 1996; Kowalski, 2002], the boundary-element method (BEM) [Stenroos and Nenonen, 2012], and the method of auxiliary sources (MAS) [Kaklamani and Anastassiu, 2002] are commonly used to simulate eddy current distributions. The FEM is particularly popular as it is suitable for objects with any shape and material properties, especially in non-destructive testing (NDT) applications [Lu et al., 2016; Lu et al., 2018]. In theory, the basic process of the FEM is to replace an entire continuous domain by a number of meshed sub-domains. The problem then becomes one of solving a system of algebraic equations, and its result is a numerical solution of the boundary problem [Reddy, 1993].

Currently, the two main methods for accelerating eddy-current calculations focus on the electromagnetic (EM) formulation strategy and on the numerical solution process. They each have advantages and disadvantages. Thus formulation-strategy methods are based on novel model decompositions employing routines such as ParaFEM [Lenzi et al., 2013], multi-layered conductive structures (MCS) [Stuchly and Esselle, 1992; Alekseev and Ziskin, 2009], and second-order transmission conditions (SOTC) [Xue and Jin, 2014]. ParaFEM is a portable library of parallel finite element analysis routines that can be used for large finite-element problems, but there is a significant reliance on the model's material distribution. The principle is to decompose the model into several continuous subdomains and then solve these subdomains in parallel; but this method assumes weak coupling between the subdomains, especially between adjacent subdomains. MCS methods have been found to be much faster than conventional FEM (by nearly 100 times), but they are restricted to certain layer-isotropic models such as in airplanes and pipeline components rather than structures with arbitrary geometry or material distribution models with the flawed structure or with mixed materials. As promotion of Finite Element domain decomposition method (DDM) [Farhat and Roux, 1992; Ma and Jin, 2014; Jin, 2002; Li and Jin, 2007] in the non-conformal (NC) framework, Second-order transmission condition (SOTC) method has been found to be more accurate on iterations over 200 but with little improvement on computation with iterations smaller than 50. Moreover, this method was also proved to be weak relationships between subdomains. FEM-BEM hybrid method [Hertel, 2007;

Matsuoka and Kameari, 1988; Navon et al., 1983; Meyer et al., 2003; Gençer et al., 2003; Kojima et al., 2001; Eibert et al., 1999] basically couples the boundary-element region with the Finite Element region to derive solutions for uniform material distributions. This method was well used in most cases but cannot be applied to several separated models that are very far from each other.

By contrast routines such as SuiteSparse [Georgii and Westermann, 2010] and GRID [Fritschy et al., 2005] speeding up the solution process for the linear equations derived from FEM. In addition, the introduction of GPUs [Maimaitijiang et.al., 2011; Xu et.al., 2010; Guan et.al., 2014; Yu and Li, 2015; Dawson et.al., 1997; Brown and Seagar,1987] can accelerate calculations by parallelisation, but at a potential increase in cost.

A different approach is taken in this work. A custom FEM solver based on edge elements is presented in which weak coupling theory based on Galerkin formulation [Bíró, 1999; Belytschko et al., 1994]. The weak coupling theory is used to reduce the vector eddy current equation to a scalar one. With this solver, eddy currents can be computed, for example, in complex biological tissue. As an illustration, the eddy current distribution is obtained in a 2M-elements realistic human brain model under normal and pathological conditions. Typical FEM solvers cannot be applied owing to the model with a large number of elements.

## Methods of Edge-Element FEM Solver

In this section, the canonical edge-element FEM solver- Galerkin formulations based FEM solver is presented. Galerkin's equations were formulated over 10 decades ago and are still the most common solver for calculating the global stiffness matrix in FEM [Bíró, 1999; Belytschko et al., 1994]. First, the original Galerkin's equations applied in electromagnetics (EM) are transformed into matrix form, as shown in the following:

$$\int_{\Omega_c} \nabla \times \mathbf{N}_i \cdot v \nabla \times \mathbf{A}^{(n)} d\Omega + \int_{\Omega_c} j\omega\sigma\mathbf{N}_i \cdot \mathbf{A}^{(n)} d\Omega + \int_{\Omega_c} j\omega\sigma\mathbf{N}_i \cdot \nabla V^{(n)} d\Omega = \int_{\Omega_c} \nabla \times \mathbf{N}_i \cdot v_0 \nabla \times \mathbf{A}_s d\Omega \quad (1)$$
$$i = 1, 2, ..., 6$$

$$\int_{\Omega_c} j\omega\sigma\nabla L_i \cdot \mathbf{A}^{(n)} d\Omega + \int_{\Omega_c} j\omega\sigma\nabla L_i \cdot \nabla V^{(n)} d\Omega = 0 \quad (2)$$
$$i = 1, 2, ..., 4$$

where $L_i$ denotes the elemental interpolation of $i^{th}$ node corresponding to its nth element, $N_i$ the vector interpolation of the $i^{th}$ edge corresponding to its nth edge element, $A^{(n)}$ the estimated edge vector potential of the nth element, V(n) the electrical potential on the 4 vertices for the nth element in the meshed domain (including the tested plate subdomains and the surrounding air subdomains), $A_s$ the original edge vector potential of the nth element, $v$ the reluctivity (the reciprocal of the permeability) of the target, $\sigma$ the conductivity of the target, and $v_0$ the reluctivity (the reciprocal of the permeability) of the air. $\nabla$ indicates the gradient calculation. $\nabla \times$ denotes the curl degree computation.

Assume for an arbitrary element *n*, there exists a matrix Q that can represent the stiffness matrix form of equation (1) and (2) left side for an arbitrary element:

$$Q^{(n)} = \begin{pmatrix} K^{(n)}(6\times 6) & L^{(n)}(6\times 4) \\ M^{(n)}(4\times 6) & N^{(n)}(4\times 4) \end{pmatrix} \tag{3}$$

Combining equation (3) with equation (1) and (2), the Galerkin matrices form of whole elements mesh becomes,

$$\sum_{n=1}^{N0} Q^{(n)} \cdot \begin{pmatrix} A^{(n)} \\ V^{(n)} \end{pmatrix} = \sum_{n=1}^{N0} \begin{pmatrix} K^{(n)} & L^{(n)} \\ M^{(n)} & N^{(n)} \end{pmatrix} \cdot \begin{pmatrix} A^{(n)} \\ V^{(n)} \end{pmatrix} = B \tag{4}$$

Here the matrix K is divided into matrices $K_1$ and $K_2$. $K_1$ denotes the matrix form of first vector potential related term in equation (1), which acts as a fundamental form of the vector potential. $K_2$ denotes the matrix form of second vector potential related term in equation (1), which exhibits the skin effect of the eddy current. L is the matrix form of the first electric potential related term in equation (1), which controls the eddy current by the Maxwell-Wagner effect- restricting the current by the shape of the target. M and N are the matrices form of the first and second term in equation (2), which collectively controls the magnetostatic field part. B is the matrix form of the right side term of equation (1) and (2), which represents the Dirichlet Boundary Condition. N0 denote the number of whole mesh elements.

The electric field of an arbitrary element can also be derived from the vector potential and electric potential in equation (4) by using the derivative of traditional **A** and **V-A** formulation incorporating Coulomb gauge [Zeng et al., 2009]:

$$E^{(n)} = -j\omega A^{(n)} - j\omega \nabla V^{(n)} \tag{5}$$

where $\mathbf{A}(n)$ denotes the vector sum of the vector potential on all the edges of each tetrahedral element, and $V(n)$ denotes the electric potential on all the vertices of each tetrahedral element.

Thus, the transmitter-receiver mutual inductance changes associated with the given model can also be obtained by applying the equation presented by Mortarelli [1980] or Auld and Moulder [1999]. In both analyses, the authors start from the Lorentz reciprocity relation and arrive at the same generalized equation that can be applied to any pair of coils [Mortarelli, 1980; Auld and Moulder, 1999]:

$$\Delta L = \frac{1}{j\omega I^2} \int_c \mathbf{E}_a \cdot \mathbf{J}_b dv = \frac{1}{j\omega I^2} \int_c \mathbf{E}_a \cdot \mathbf{E}_b \cdot (\sigma_a - \sigma_b) dv \qquad (6)$$

Here, $\Delta L$ denotes the inductance changes caused by the difference between medium a and b.

# Improved Method

## A. Weakly coupled approximation method

To develop an efficient FEM solver that can simulate the eddy currents induced by an excitation coil in biological tissues in any sectional views for normal and pathological conditions (such as internal and peripheral strokes), a custom FEM formulation based on the Rayleigh-Ritz-Galerkin method under low frequency is presented in equation (8).

Since the value of the current angular frequency $\omega$ is small under the low-frequency condition, the value of the $\int_{\Omega_c} j\omega\sigma \mathbf{N}_i \cdot \mathbf{A}^{(n)} d\Omega$ component is much smaller than that of $\int_{\Omega_c} \nabla \times \mathbf{N}_i \cdot v \nabla \times \mathbf{A}^{(n)} d\Omega$, which suggests the secondary field produced by the eddy currents are very small compared to the primary field. This approximation is called the weakly coupled approximation. Then, ignoring the $\omega$ terms, equation (1) becomes,

$$\int_{\Omega_c} \nabla \times \mathbf{N}_i \cdot v \nabla \times \mathbf{A}^{(n)} d\Omega \approx \int_{\Omega_c} \nabla \times \mathbf{N}_i \cdot v_0 \nabla \times \mathbf{A}_s d\Omega$$
$$i = 1, 2, ..., 6 \qquad (7)$$

Assuming the reluctivity of the target $v$ is approximately equal to that in the air $v_0$, the estimated edge vector potential $\mathbf{A}^{(n)}$ is approximately the same as the original edge vector potential. Therefore,

$\mathbf{A}^{(n)} \approx \mathbf{A}_s$. Consequently, the electric potential on all the vertices of each tetrahedral element can be calculated by the following:

$$\int_{\Omega_c} j\omega\sigma\nabla L_i \cdot \mathbf{A}_s d\Omega + \int_{\Omega_c} j\omega\sigma\nabla L_i \cdot \nabla V^{(n)} d\Omega \approx 0$$
$$i = 1, 2, ..., 4 \tag{8}$$

And the boundary condition $\mathbf{A}_s$ can be linked to the current flowing transmitter based on the Biot-Savart law [Hayashi et al., 1989].

Then the electric field can be obtained once V is known:

$$\mathbf{E}^{(n)} = -j\omega \mathbf{A}_s^{(n)} - j\omega\nabla V^{(n)} \tag{9}$$

Since only scalar electric field $\nabla V^{(n)}$ remains in equation (8) under low frequency, the eddy current computation time will be much reduced when compared to vector electric field in equation (5).

Compared with the typical FEM solver (equation 1 and 2), the proposed simplified FEM solver (equation 8) can significantly reduce the computation burden as only one polynomial needs to be solved. In addition, the proposed simplified FEM solver has eliminated the frequency terms in typical FEM solver (equation 1), which would require a larger region to be defined to encompass the original model in order to precisely compute the eddy currents affected by the skin effect. Therefore, the overall mesh size is much larger than 2 million when calculated by the typical FEM solver.

Based on the simplified FEM algorithm (equation 8), the Sparse Reverse Cuthill-McKee orderings technique and the incomplete LU (ILU) decomposition method have been implemented in the proposed FEM solver to reorder and decompose the sparse polynomial matrix ( $j\omega\sigma \text{grad} L_i$ term in equation 8) so that most of the elements will be distributed close to the diagonal of the stiffness matrix (the sparse polynomial matrix), which was showed to be very efficient in accelerating the convergence of the solving process [Bridson and Tang, 2001]. The 2M realistic human brain model's stiffness sparse matrix elements distributions before and after the reordering and decomposing are presented in Figure 1 a) and b). After reordering and decomposing the sparse stiffness matrix, Bi-conjugate Gradients Stabilised (CGS) iterative method was utilized to find the final numerical solution of the system.

To validate the proposed FEM solver, numerical simulations were performed by our solver and a commercial solver (COMSOL) over a cylindrical rod model with various diameters and offsets. For the same set of simulations, we kept the model geometries the same for our solver and COMSOL solver so that the comparison is meaningful. In addition, the mesh file extracted from the COMSOL was used for our solver, so that the meshes for the object(s) under our solver simulation were exactly the same.

Firstly, we simulated a simple setup where a cylindrical object with height 1 m was placed axial-symmetrically between two co-axial coils (Fig. 2). The relative permeability and relative permittivity of this object were set to be 1 and its conductivity is 1 S/m. The distance between two coils was fixed at 3 m. The exciting frequency was 1 kHz in order to obey the weakly coupled approximation.

For the cylindrical objects of different radius, the changes in the induced voltage due to the presence of the object are shown in Table 1.

For the coils of different radius, the changes in the induced voltage due to the presence of the object are shown in Table 2.

The results in Table 1and Table 2 present high coherence between our solver and the COMSOL solver under different setups.

Furthermore, we tested the case where the cylindrical objects are positioned off the axis. Due to the break of axial symmetry, 3D models have to be used in COMSOL while our solver is inherently 3D. So only the position of the object was moved to simulate this setup. Table 3 summarises the results.

The discrepancy between two solvers increased when the object moved far away off the symmetric axis. As the object moves away from the axis, the charge accumulated on the boundary of the object increases and the contribution to E from electrical potential V term becomes significant.

Lastly, in order to compare the speed of the two solvers, Table 4 summarises the solution time for specific setups over the frequencies at 1 kHz, 5 kHz, 10 kHz, 50 kHz, 100 kHz, 300 kHz, 500 kHz, 800 kHz, and 1 MHz.

From Table 4, our solver is faster in solving FEM electromagnetic models than the COMSOL solver for using the weakly coupled assumption. Eddy current distributions from our solver are also plotted and they follow the same pattern in COMSOL (Table 5). In our previous work, the weakly coupled approximation has been further validated by simulating the magnetic induction tomography in another realistic head model [Dekdouk et al., 2010].

Once our Edge-element FEM solver was validated by comparing the accuracy and efficiency with COMSOL, the eddy current distribution within the 2M realistic human brain can also be calculated by the proposed Edge-element FEM simulation.

## B. Models

In this paper, a 2 million-element realistic human brain with a dimension of 172x211x192mm is simulated (Fig. 3) under both normal and pathological conditions (internal and peripheral strokes). The mesh statistics of the realistic human brain mesh model(University College London, London, UK) is presented in Table 6 [Tizzard et al., 2005; Yerworth et al., 2003; Tidswell et al., 2001; Holder et al., 2003; Romsauerova et al., 2006]. In addition, the conductivity of brain tissues is listed in Table 7. The conductivity of pathological stroke equals 25% of tissue (grey or white matter) conductivity plus 75% of human brain blood conductivity. Here, it is necessary to mention that the electrical properties (conductivity) of human brain blood depend on various factors such as cell volume fractions and viscosity [Beving et al., 1994; Abdalla, 2011]. However, the average value of human brain blood electrical conductivity is measured to be 1.818 m S/m [Turner, 1902]. As for realistic stroke data, it would be difficult for us at this stage to obtain realistic data for the stroke cases, as it is a hugely complex situation depending on patients, circumstances, and timings, so we would not be able to simulate this. But certainly, this would be a future direction for us.

## C. Sensor configuration

The brain model is excited by a coil of 62.5 mm in radius with a current of 1 A under a frequency of 10 Hz, positioned on its top with a distance of 147 mm to the brain center. A receiver coil is also located on its top with a distance of 145 mm to the brain center. The sensor setup and parameters (both transmitter and receiver) are shown in Figure 4 and Table 8. The sensor is applied for all the simulations from Figure 3 to Figure 11.

## D. Results

Using the proposed improvement method in part A, the eddy current simulations of 2 million elements of the realistic brain model under the normal and pathological (internal and peripheral strokes) condition in low frequency (here is 10 Hz) now become possible, which cannot be solved by typical commercial

FEM solvers. Note that in order to solve this problem using a commercial solver, a larger region needs to be defined such that the overall mesh size is much larger than 2 million.

**D.1 Eddy current simulations of brain model under normal condition**

In Figure 5, the model is excited by a transmitter above with an alternating current flow of 1 Ampere under 10 Hz (dimensions and locations are presented in part C). As can be seen from the eddy current plots and their legends, the solver is validated to be accurate since the eddy current flows smoothly and continuously within each brain tissue (with higher eddy current density in CSF and lower eddy current density in air cavities). And a higher eddy current density on the top of the brain is due to the above-located transmitter as shown in part C.

**D.2 Eddy current simulations of brain model under internal stroke condition**

In Figure 6, the pathological stroke is only encountered within the white and grey matter.

For Figure 7, the location and properties of probes are the same as that in part D1 case. And this Figure is used to compare the eddy current distributions under the internal stroke condition as in Figure 8.

In Figure 8, the location and properties of probes are also the same as that in part D1 case. By comparing the eddy current plots and their legends of Figure 7 and Figure 8, the solver is further proved to be accurate since the eddy current density on the internal stroke position (Fig. 6) is much higher than that on the same position under the normal condition. This is due to the larger conductivity of the stroke position as the conductivity of pathological stroke equals 25% of tissue (grey or white matter) conductivity plus 75% of human brain blood conductivity.

**D.3 Eddy current simulations of brain model under peripheral stroke condition**

In Figure 9, the pathological stroke is only encountered within the white and grey matter.

For Figure 10, the location and properties of probes are the same as that in part D1 case. And this Figure is used to easily compare the eddy current distributions under the peripheral stroke condition as in Figure 11.

In Figure 11, the location and properties of probes are also the same as that in part D1 case. By comparing the eddy current plots and their legends of Figure 10 and Figure 11, the solver is further validated to be accurate since the eddy current density on the peripheral stroke position (Fig. 9) is much higher than that on the same position under the normal condition. This is still due to the larger conductivity of the stroke position as the conductivity of pathological stroke equals 25% of tissue (grey or white matter) conductivity plus 75% of human brain blood conductivity. And eddy current shows to be discontinuous on the boundary of stroke.

**D.4 Comparison between canonical FEM solver and the custom built FEM solver under low frequency (10 Hz)**

We have also made the eddy current simulations for the same brain model but different mesh elements size (from 50k to 1.1M) by both canonical FEM solver (As shown in equation 1, 2)and the custom built FEM solver (As shown in equation 8 and 9)under the frequency of 10 Hz. And the computation time cost by both solvers is listed in Table 9. Here, the simulations were computed on ThinkStation(Lenovo Group Ltd, Beijing, China) P510 platform with Dual Intel Xeon E5-2600 v4 Processor, with 16G RAM. As can be seen from Table 9, the efficiency of the custom-built FEM solver is much higher than that of the canonical FEM solver (equation 1 and 2). And the simulation of brain model with a mesh elements size of 900k and 1.1M cannot be calculated by the canonical FEM solver, which is due to a too large stiffness matrix caused by the large mesh elements size when calculated by the canonical FEM solver (As shown in equation 1, 2, and 5). For instance, if a sample meshes into $m$ tetrahedral elements with $n$ total nodes and $k$ total edges, then workstation need to solve a polynomial with $(n+k) \times (k+n)$ size stiffness matrix for the typical FEM solver (as shown in equation 4). However, for the proposed FEM solver, workstation only needs to solve a polynomial with $n \times n$ size stiffness matrix (as shown in equation 8).

# Conclusions

In this paper, a custom FEM solver based on edge elements has been developed using weakly coupled theory. This solver is more efficient than typical commercial solvers since it reduces the vector eddy current equation to a scalar one, and reduces the meshing domain to just the eddy current region. Consequently, it can tackle complex eddy current calculations for models with a much larger number of elements such as those encountered in eddy current computation in biological tissue. An example of eddy currents simulation in biological tissues of a realistic human brain mesh (2 million elements) is shown. Sectional views in the x-axis, y-axis, as well as z-axis, are easily obtained for the distribution of eddy currents in all tissue types with various dielectric properties under both normal and pathological conditions. As can be seen from the results, the solver has been validated to be accurate as eddy current flows rationally (smoothly and continuously) within brain tissue under both normal and pathological condition. In addition, with this solver, the equivalent magnetic field induced from the excitation coil is applied and therefore there is no need to mesh the excitation coil, which has significantly reduced the element number and released the calculation burden. However, the proposed weakly coupled based improved FEM solver will be invalid in high-frequency computation. Therefore, a high-frequency improved FEM solver is worth considering in the future. Moreover, another promising field could be diagnosing human health condition or even detecting the pathological region position from the complex impedance data of human brain calculated by the proposed FEM solver.

## Acknowledgments

The authors would like to thank the UK Engineering and Physical Sciences Research Council (EPSRC). The authors would like to thank Professor David Holder for providing the brain mesh.

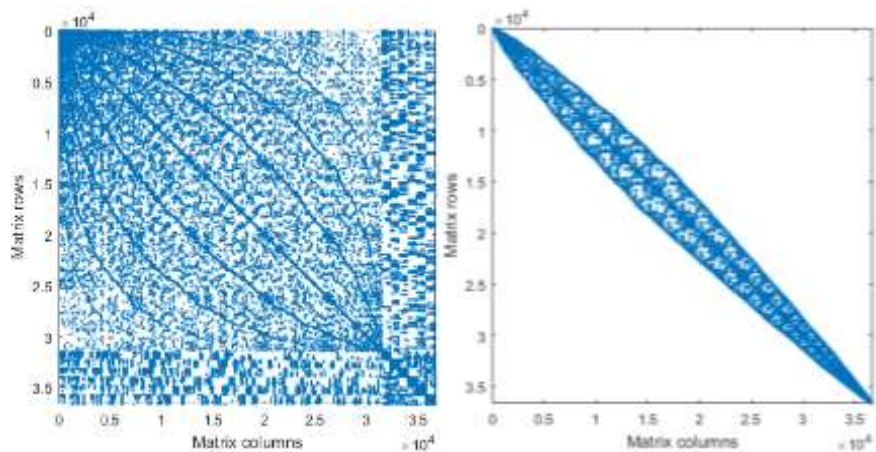

**Fig. 1. 2M realistic human brain model's stiffness sparse matrix elements distributions a) Before re-orderings and decomposition b) After re-orderings and decomposition**

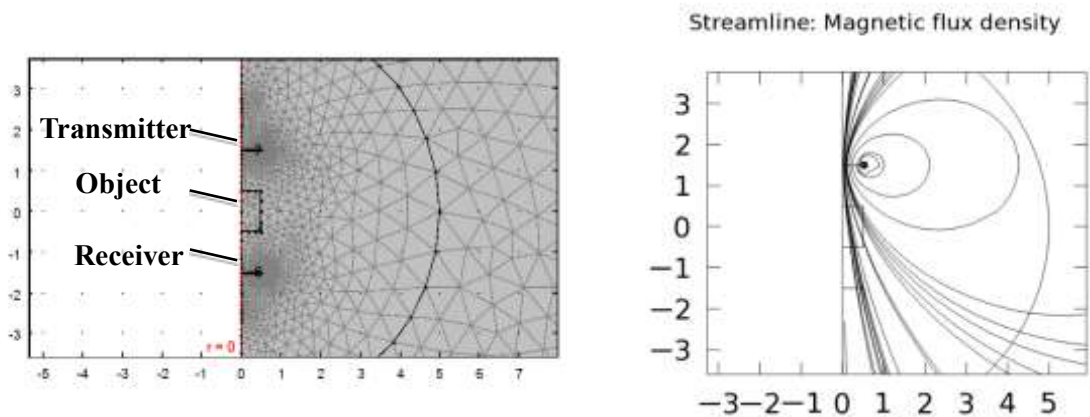

a) Sensor and object location          b) Cylindrical objects with different radius

**Fig. 2. 2D axis symmetric FEM model in COMSOL a) Sensor and object location b) Cylindrical objects with different radius**

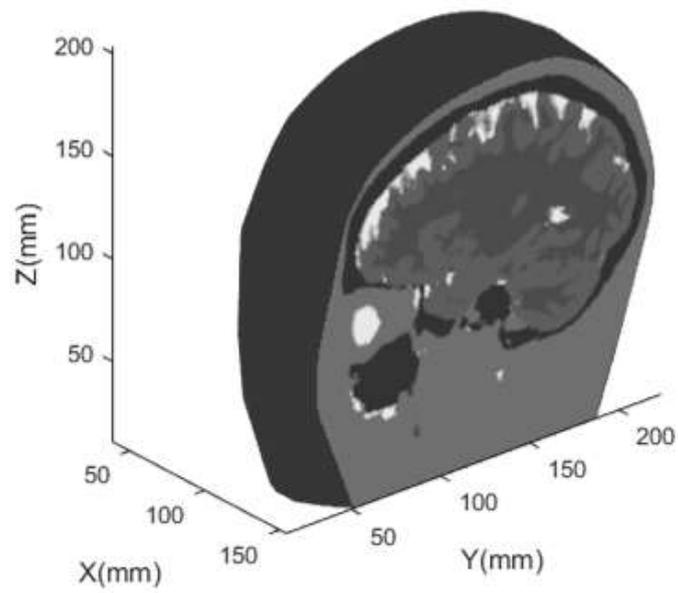

**Fig. 3. 2M elements realistic human brain model**

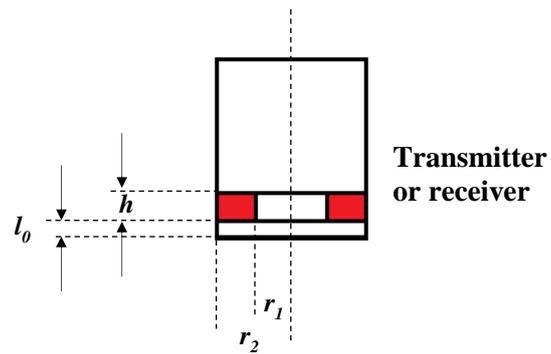

**Fig. 4. Sensor configuration**

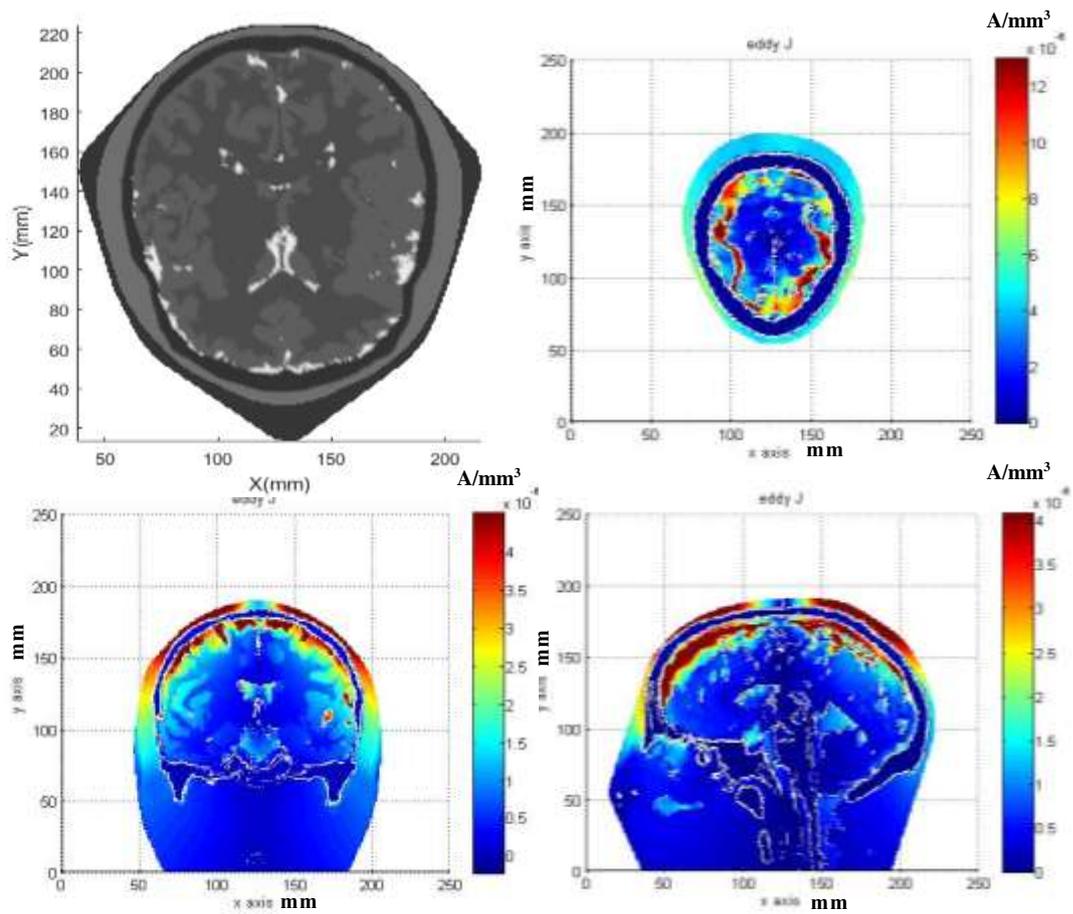

**Fig. 5. Eddy current simulations for different dimension views in a 2M elements brain model under normal condition**

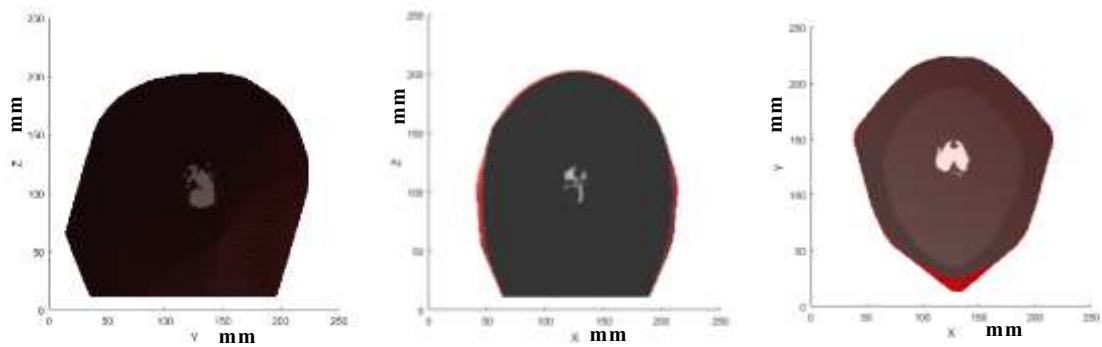

**Fig. 6. The position of internal stroke for different dimension views in the 2M elements brain model**

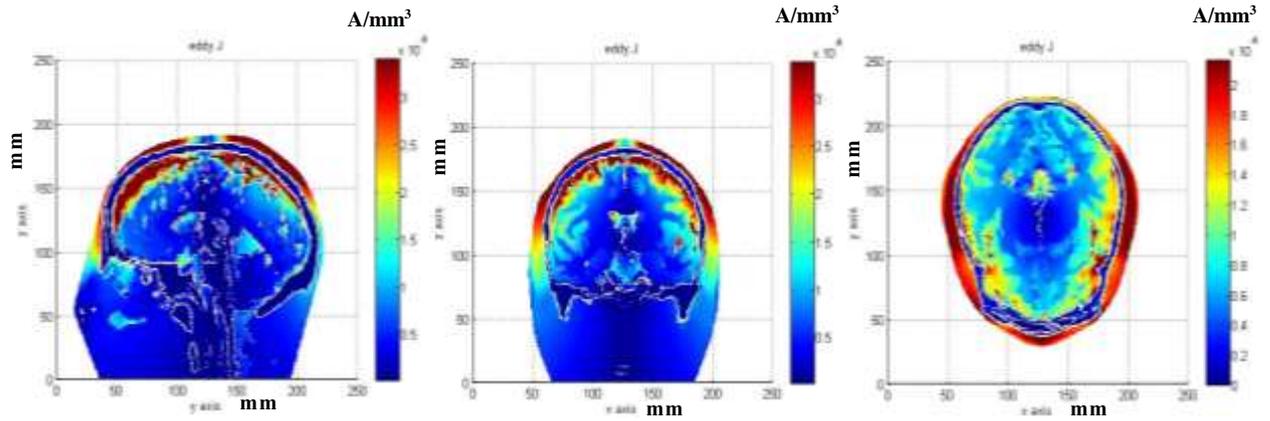

**Fig. 7. Eddy current simulations for different dimension views in a 2M elements brain model under normal condition**

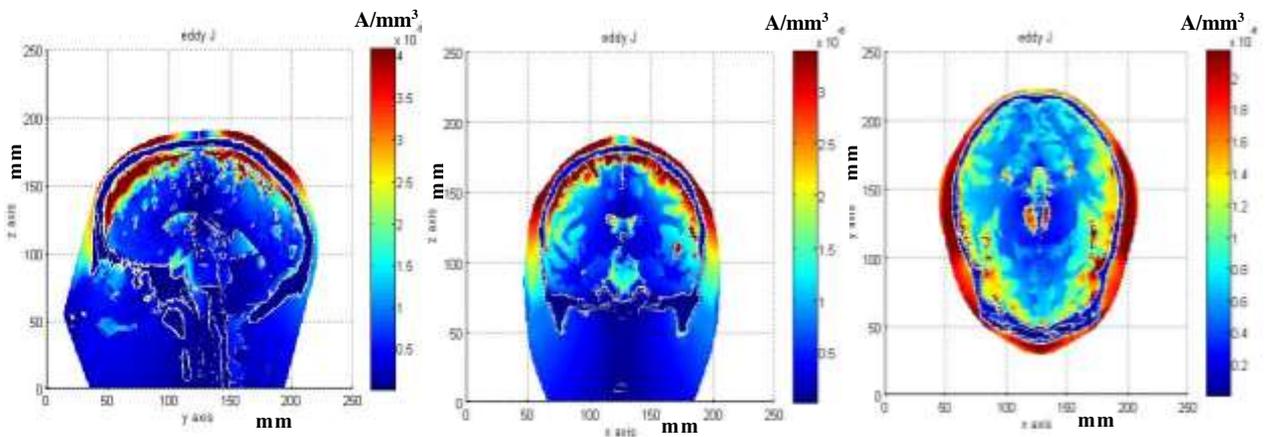

**Fig. 8. Eddy current simulations for different dimension views in a 2M elements brain model under internal stroke condition**

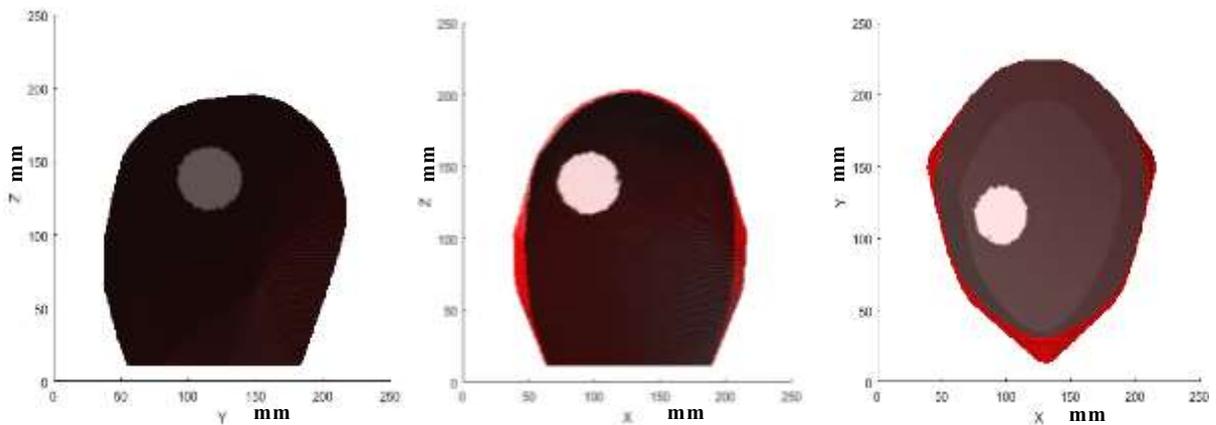

**Fig. 9. The position of peripheral stroke for different dimension views in the 2M elements brain model**

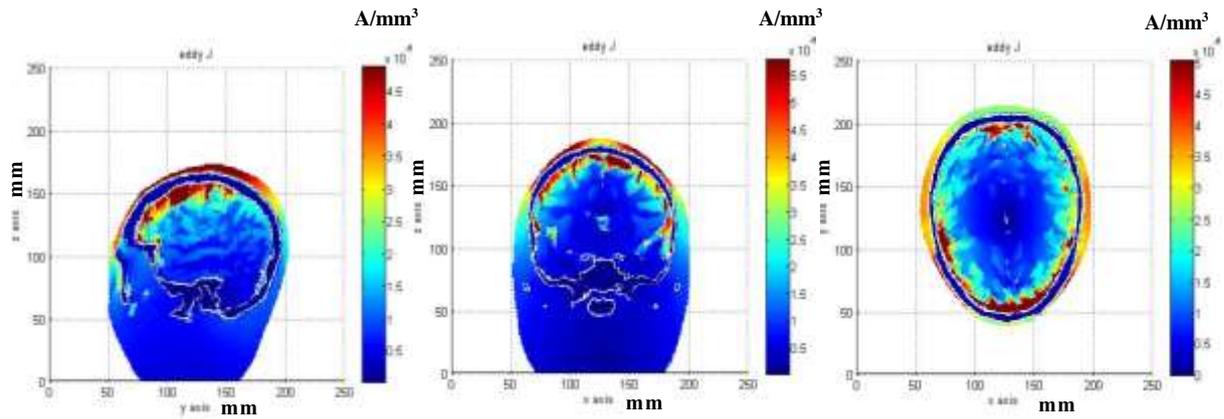

**Fig. 10. Eddy current simulations for different dimension views in a 2M elements brain model under normal condition**

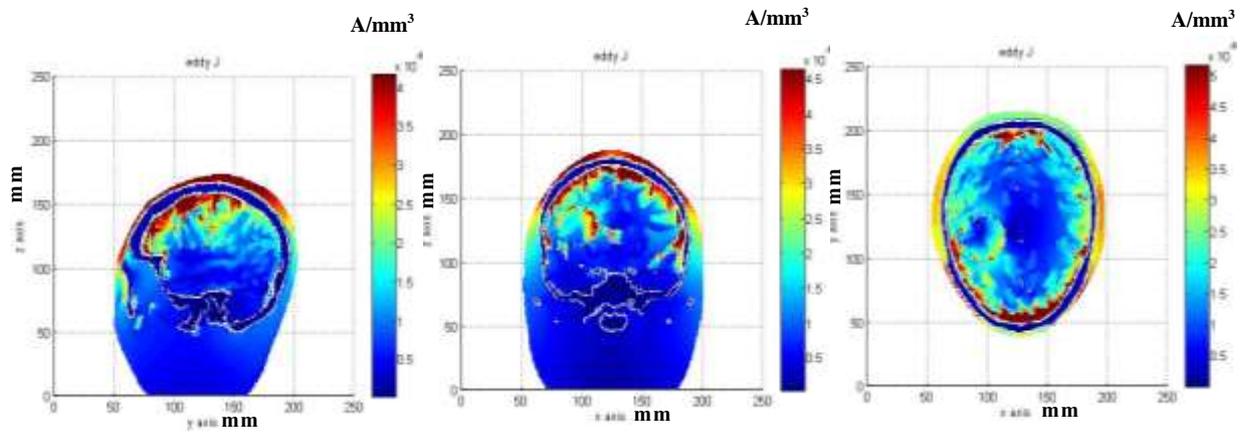

**Fig. 11. Eddy current simulations for different dimension views in a 2M elements brain model under peripheral stroke condition**

**Table 1. Induced voltage changes due to the presence of the object at the different radius (Coil radius=0.5 m)**

| The radius of the Object (m) | Our Solver (Resistive part of the impedance normalised by frequency) (ohm) | COMSOL Solver (Resistive part of the impedance normalised by frequency) (ohm) | Error (%) |
|---|---|---|---|
| 0.6 | -2.66E-12 | -2.6650E-12 | -0.19% |
| 0.5 | -1.38E-12 | -1.3853E-12 | -0.38% |
| 0.4 | -6.04E-13 | -6.0442E-13 | -0.07% |
| 0.3 | -2.01E-13 | -2.0104E-13 | -0.02% |
| 0.2 | -4.12E-14 | -4.1198E-14 | 0.00% |
| 0.1 | -2.63E-15 | -2.6337E-15 | -0.14% |

**Table 2. Induced voltage changes due to the presence of the object by different coil radius (Object radius=0.5 m)**

| The radius of the Coils (m) | Our Solver (Resistive part of the impedance normalised by frequency) (ohm) | COMSOL Solver (Resistive part of the impedance normalised by frequency) (ohm) | Error (%) |
|---|---|---|---|
| 0.6 | -2.55E-12 | -2.5435E-12 | 0.26% |
| 0.5 | -1.38E-12 | -1.3853E-12 | -0.38% |
| 0.4 | -6.31E-13 | -6.2882E-13 | 0.35% |
| 0.3 | -2.17E-13 | -2.1581E-13 | 0.55% |
| 0.2 | -4.55E-14 | -4.5135E-14 | 0.81% |
| 0.1 | -2.95E-15 | -2.8971E-15 | 1.83% |

**Table 3. Induced voltage changes due to the presence of the object at different offset (Object radius=0.5 m, coil radius=0.5 m)**

| The offset of the Object (m) | Our Solver (Resistive part of the impedance normalised by frequency) (ohm) | COMSOL Solver (Resistive part of the impedance normalised by frequency) (ohm) | Error (%) |
|---|---|---|---|
| 0.6 | -4.61E-13 | -4.8739E-13 | -5.41% |
| 0.5 | -6.68E-13 | -6.8605E-13 | -2.63% |
| 0.4 | -8.82E-13 | -8.8893E-13 | -0.78% |
| 0.3 | -1.08E-12 | -1.0824E-12 | -0.22% |
| 0.2 | -1.24E-12 | -1.2389E-12 | 0.09% |
| 0.1 | -1.35E-12 | -1.3392E-12 | 0.81% |

**Table 4. Comparison of the solution time for two solvers**

| Model | Our Solver | COMSOL Solver |
|---|---|---|
| (a). coil radius =0.5 m; object radius =0.5 m; offset =0 m | 4.8 s | 513 s |
| (b). coil radius =0.5 m; object radius =0.5 m; offset =0.2 m | 4.8 s | 445 s |
| (c). coil radius =0.5 m; object radius =0.5 m; offset =0.3 m | 4.8 s | 544 s |
| (d). coil radius =0.5 m; object radius =0.5 m; offset =0.4 m | 4.8 s | 475 s |
| (e). coil radius =0.2 m; object radius =0.5 m; offset =0 m | 4.8 s | 393 s |
| (f). coil radius =0.2 m; object radius =0.5 m; offset =0.2 m | 4.8 s | 491 s |

**Table 5. Eddy current patterns from our solver**

| | |
|---|---|
| (a) coil radius =0.5 m; object radius =0.5 m; offset =0 m | 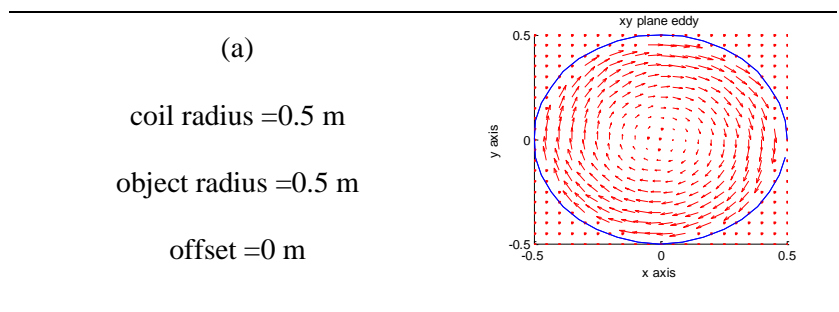 |

| | | |
|---|---|---|
| (b) coil radius =0.5 m  object radius =0.5 m  offset =0.2 m | 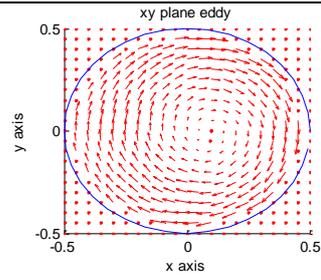 | |
| (c) coil radius =0.5 m  object radius =0.5 m  offset =0.3 m | 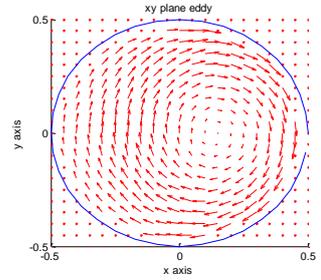 | |
| (d) coil radius =0.5 m  object radius =0.5 m  offset =0.4 m | 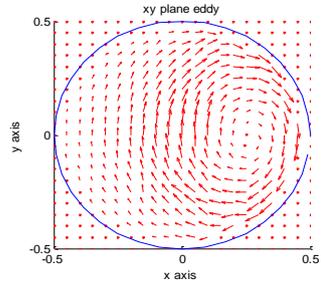 | |
| (e) coil radius =0.2 m  object radius =0.5 m  offset =0 m | 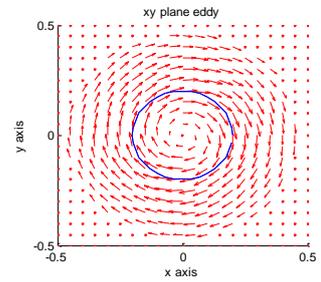 | |
| (f) coil radius =0.2 m  object radius =0.5 m  offset =0.2 m | 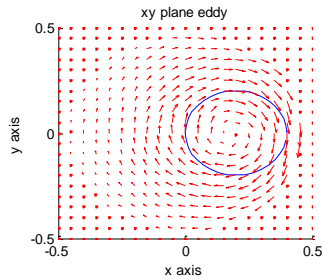 | |

**Table 6. Mesh statistics for 2M elements realistic human brain model**

| Number of elements | 2,047,408 |
|---|---|
| Number of nodes | 349,305 |

| Minimum intermodal spacing | 0.9801 E-6 mm |
|---|---|
| Average intermodal spacing | 0.5094 mm |

**Table 7. The electrical properties of brain tissues**

| Tissues | Conductivity (S/m) |
|---|---|
| white matter | 0.15 |
| grey matter | 0.3 |
| CSF | 1.79 |
| dura mater | 0.44 |
| skull | 0.018 |
| air cavities | 0.001 |
| scalp | 0.44 |

**Table 8. Coil parameters**

| $r_1$ | 62.5mm |
|---|---|
| $r_2$ | 65mm |
| lo (lift-off) | 2mm |
| h (height) | 2.2mm |
| Number of turns $N_1$(Transmitter) = $N_2$(receiver) | 80 |

**Table 9. Comparison between Canonical FEM solver and the custom built FEM solver under low frequency (Here is 10 Hz)**

|  | Computation Time (s) |
|---|---|

| Brain model mesh size | Canonical FEM solver (equation 1 and 2) | The custom built FEM solver |
| --- | --- | --- |
| ~50k | 426 | 68 |
| ~100k | 1162 | 167 |
| ~300k | 3749 | 435 |
| ~500k | 9758 | 1083 |
| ~700k | 14853 | 1394 |
| ~900k | Infeasible | 1652 |
| ~1.1M | Infeasible | 1978 |